\documentclass[aps,twocolumn,prl,amsmath,amssymb,superscriptaddress]{revtex4}
\usepackage{multirow}
\usepackage{graphicx,epsfig}
\usepackage{wasysym}

\begin{document}
\title{Dynamic thermoelectric and heat transport in mesoscopic capacitors}
\author{Jong Soo Lim}
\affiliation{Instituto de F\'{\i}sica Interdisciplinar y Sistemas Complejos
IFISC (UIB-CSIC), E-07122 Palma de Mallorca, Spain}
\affiliation{School of Physics, Korea Institute for Advanced Study,
Seoul 130-722, Korea}
\author{Rosa L\'opez}
\affiliation{Instituto de F\'{\i}sica Interdisciplinar y Sistemas Complejos
IFISC (UIB-CSIC), E-07122 Palma de Mallorca, Spain}
\affiliation{Departament de F\'{\i}sica,
Universitat de les Illes Balears, E-07122 Palma de Mallorca, Spain}
\author{David S\'anchez}
\affiliation{Instituto de F\'{\i}sica Interdisciplinar y Sistemas Complejos
IFISC (UIB-CSIC), E-07122 Palma de Mallorca, Spain}
\affiliation{Departament de F\'{\i}sica,
Universitat de les Illes Balears, E-07122 Palma de Mallorca, Spain}

\begin{abstract}
We discuss the low-frequency response of charge and heat transport
to oscillatory voltage and temperature shifts in mesoscopic capacitors.
We obtain within scattering theory generic expressions for the quantum admittances
up to second order in the ac frequencies in terms of electric, thermoelectric
and heat capacitances and relaxation resistances.
Remarkably, we find that the thermocurrent can lead or lag
the applied temperature depending on the gate voltage
applied to a quantum $RC$ circuit. Furthermore, the relaxation
resistance for cross terms becomes nonuniversal as opposed
to the purely electric or thermal cases.
\end{abstract}


\maketitle

\paragraph{Introduction.}Time-dependent
charge transport in quantum conductors
subject to ac electric fields provides insight
into electronic dynamics at nanoscale dimensions \cite{but93a}.
For low frequencies the dynamics of a mesoscopic conductor
is characterized by the $R_GC_G$ time,
where $C_G$ is the quantum capacitance and $R_G$ is 
the charge relaxation resistance \cite{but93b}.
Whereas the former generally depends on energy, the latter
unexpectedly attains the constant value of $h/2e^2$
in a single-channel quantum capacitor \cite{but93a}.
These predictions are confirmed by ac measurements
in mesoscopic $RC$ circuits~\cite{gab06}.
Further developments have led to the experimental demonstration
of coherent single-electron emitters~\cite{feb07}.
This achievement has spurred an enormous interest
in both the fundamentals of time-resolved electronic transport
(both experimentally~\cite{chorley12,basset12,frey12}
and theoretically~\cite{nig06,wan07,rin08,mos08,mor10,cot11,lee11,bat13})
and its applications to, e.g., metrology~\cite{pek12}
and quantum information processing~\cite{spl09},
just to mention a few.

Electronic current, however, can also be driven by thermal gradients.
In the stationary case the Seebeck effect leads to the generation
of thermovoltages in response to applied temperature differences
in open circuits~\cite{book}. While dc thermopower has been extensively
investigated in nanostructures~\cite{mol92,dzu97,red07,sch08,mat12},
the ac Seebeck effect has received little attention to date.
The subject is interesting
for several reasons. First, treating voltage and thermal driving
fields on an equal footing opens up the door to 
not only electrical but also thermodynamic characterizations of mesoscopic systems.
In fact, ac calorimetry techniques have been successfully applied
to superconducting loops~\cite{bou05}. What can be learned from an analogous
experiment with normal conductors? Second,
quantum refrigeration devices based on the Peltier effect (reciprocal to Seebeck)
typically use static currents~\cite{gia06}. What can we expect in the ac regime
of transport? Third, how does Coulomb interaction 
renormalize the $RC$ parameters in a thermoelectric device?
These are the kind of questions we want to address in this work.
\begin{figure}[!h]
\begin{center}
\includegraphics[width=0.45\textwidth]{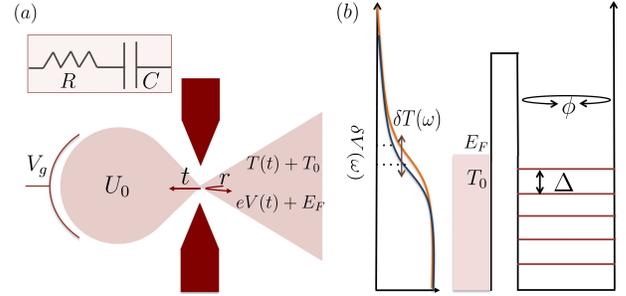}
\caption{(a) Sketch of a mesoscopic capacitor: a large dot attached to a reservoir
with time-dependent (oscillatory) temperature $T(t)$ and voltage $V(t)$
with respect to the base temperature $T_0$ and chemical potential $\mu\simeq E_F$.
The coupling region is a single-mode quantum point contact
with transmission $t$ and reflection $r$
amplitudes modulated with external gates (dark shaded areas). Interactions are modelled
with a homogeneous potential $\delta U_0$ that reacts to a change in a nearby gate contact
coupled via a geometrical capacitance $C$. Electric, thermoelectric and heat properties 
of the capacitor are described at low frequency $\omega$
with an equivalent $RC$ circuit, where $R_\mathcal{A}$ and $C_\mathcal{A}$ depends on the transport type
(charge, energy, or both).
(b) Energy diagram of the mesoscopic capacitor. The Fermi distribution function in the reservoir
changes with the oscillatory voltages. The dot contains many levels with mean spacing $\Delta$.
$\phi$ is the electronic accumulated phase in a single turn around the dot.}
\label{figure1}
\end{center} 
\end{figure}

Consider a multichannel mesoscopic conductor coupled to a single terminal
as in Fig.~\ref{figure1}(a).
The sample is driven out of equilibrium with oscillating
voltages $\delta V(\omega)$ \cite{gab06}
and temperatures $\delta T(\omega)$ \cite{che01}
applied to the reservoir. The driving fields
operate with regard to an equilibrium state
described by the chemical potential $\mu$
and the base temperature $T_0$. Hence,
the linear-response electric $\delta I$ and heat $\delta J$
currents are given by
\begin{equation}\label{eq_matrix}
\begin{pmatrix}
\delta I\\ \delta J
\end{pmatrix}
=
\begin{pmatrix}
G(\omega)& L(\omega)\\ M(\omega)& K(\omega)
\end{pmatrix}
\begin{pmatrix}
\delta V\\ \delta T
\end{pmatrix} \,,
\end{equation}
where the $2\times 2$ Onsager matrix includes diagonal
elements [electric $G(\omega)$ and thermal $K(\omega)$
admittances] and nondiagonal coefficients
[thermoelectric $L(\omega)$ and electrothermal $M(\omega)$
admittances]. The latter are related by reciprocity \cite{ons31}.

In a low-frequency expansion,
$G(\omega)=-i\omega C_G+\omega^2 C_G^2 R_G$,
the imaginary part of the electric admittance
provides information on the electric capacitance or emittance of the system
whereas its real part is directly related to 
the dissipation in the conductor. 
When transport is phase coherent, $R_G=h/2e^2$ takes an universal
value independently on the transmission value and it scales with
$N$ the number of propagating channels \cite{but93a}.
Below, we demonstrate that the low temperature
thermal relaxation resistance also becomes quantized
but, surprisingly, its thermoelectric analog
is always nonuniversal. Importantly, our gauge-invariant
theory also includes the effect of Coulomb interactions
taking into account the charges that pile-up when the
sample is driven out of equilibrium by oscillatory voltage
and thermal biases. We will illustrate
our ac thermolectrical scattering theory
with an application to a  quantum capacitor system. 

\paragraph{Scattering formalism.}We present a scattering approach for coupled charge
and energy transport. In general, the transport properties of a multichannel mesoscopic capacitor are described by the scattering matrix $S$ with elements $s_{nm}[E,U(\vec{r})]$ that relate the outgoing current amplitude in channel $n$ to the incoming current amplitude in channel $m$ for a carrier with a given energy $E$. The matrix is also a function of the internal potential landscape $U(\vec{r})$ built up inside the conductor. Quite generally, $U(\vec{r})$ is a function of the applied electrical and thermal biases~\cite{but93,san13}. Let us for the moment focus on the noninteracting case.
Then, the time-dependent charge and heat current operators are given by  $\hat{I} = (e/h) \int dE dE' \exp[i(E-E')t/\hbar]  [a^{\dag}(E)a(E') - b^{\dag}(E)b(E')]$ and $\hat{J}  = (1/2h)\int dE dE' (E+E'-2\mu)\exp[i(E-E')t/\hbar] [a^{\dag}(E)a(E') - b^{\dag}(E)b(E')]$
where $a = (a_{ 1},a_{ 2},\cdots,a_{N})^\dagger$ and  $b = (b_{ 1},b_{ 2},\cdots,b_{N})^\dagger$
denote incoming and outgoing annihilation operators in the lead with $N$ channels
and fulfill $b_{n}(E) = \sum_{m} s_{nm}(E) a_{m}(E)$. 
We hereafter take $\mu\simeq E_F$ (the Fermi energy)
independently of $T_0$, which is a good approximation at low temperature,
the regime of our interest.

The linear response of the electrical and heat currents to the ac external perturbations in Eq.~\eqref{eq_matrix}
are expressed as $G(\omega)=\delta I(\omega)/\delta V(\omega)$,
$L(\omega)=\delta I(\omega)/\delta T(\omega)$,
$M(\omega)=\delta J(\omega)/\delta V(\omega)$, and
$K(\omega)=\delta J(\omega)/\delta T(\omega)$, where $I=\langle\hat{I}\rangle$
and $J=\langle\hat{J}\rangle$.
The admittances can be obtained from the Kubo formulas \cite{mahan,kubo,callen},
\begin{subequations} \label{eq_GKLM}
\begin{align}
G(\omega) &= \frac{1}{\hbar\omega} \int_{0}^{\infty}\! dt~ e^{i(\omega + i0^+)t} \langle [\hat I(t),\hat I(0)]\rangle\,, \label{eq_g}\\ 
K(\omega) &= \frac{1}{\hbar\omega T_0} \int_{0}^{\infty}\! dt~ e^{i(\omega + i0^+)t} \langle [\hat J(t),\hat J(0)]\rangle\,, \label{eq_k}\\ 
L(\omega) &= \frac{1}{\hbar\omega T_0} \int_{0}^{\infty}\! dt~ e^{i(\omega + i0^+)t} \langle [\hat I(t),\hat J(0)]\rangle\,, \label{eq_l}\\ 
M(\omega)&= \frac{1}{\hbar\omega} \int_{0}^{\infty}\! dt~ e^{i(\omega + i0^+)t} \langle [\hat J(t),\hat I(0)]\rangle\,.\label{eq_m}
\end{align}
\end{subequations}
By inserting the above expressions for $\hat{I}$ and $\hat{J}$ in Eq.~\eqref{eq_m},
we find the reciprocity relation $M(\omega)= T_0 L(\omega)$, as
expected. In the presence of an external magnetic field $B$, reciprocity becomes
$M(\omega,B)= T_0 L(\omega,-B)$. An important remark is now in order.
Concerns have been recently raised about the validity
of the fluctuation-dissipation theorem
applied to heat transport~\cite{ave10}. In fact,
Refs.~\cite{ave10,ser11} find a nonvanishing term for the equilibrium
heat-heat correlation function at $T_0\to 0$, which is incompatible
with the expected behavior of $K(\omega)$. However, this term
is associated to scattering events that connect {\em two} different terminals
and can therefore be safely ignored in our single-lead quantum capacitor system.

Inserting the charge and current operator expressions in Eq. (\ref{eq_GKLM}) we find
\begin{subequations} \label{eq_GKL}
\begin{align}
G(\omega) &= \frac{e^2}{h} \int dE\,
\text{Tr}\,A(E,E+\hbar\omega) \bar{F}(E,\omega)\,,\\ 
K(\omega) &= \frac{1}{hT_0} \int dE \,E_\omega^2 \,
\text{Tr} \,A(E,E+\hbar\omega) \bar{F}(E,\omega)\,, \\ 
L(\omega) &=\frac{e}{hT_0} \int dE \,E_\omega \,
\text{Tr} \,A(E,E+\hbar\omega) \bar{F}(E,\omega) \,,
\end{align}
\end{subequations}
where the trace is over the transverse channels,
$E_\omega=(E+\hbar\omega/2-E_F)$,  $A(E,E+\hbar\omega)=1- S^{\dagger}(E)S(E+\hbar\omega)$,
and $\bar{F}(E,\omega)=[f_0(E)-f_0(E+\hbar\omega)]/\hbar\omega$
with $f_0=1/(1+\exp[(E-E_F)/k_BT_0])$ the Fermi function of the equilibrium reference state.

\paragraph{Low frequency analysis.}The quantum capacitor exhibits pure ac response.
Then, we expand Eq.~\eqref{eq_GKL} in powers of $\omega$.
The leading-order contributions have the following functional form for $\mathcal{A}=G$, $K$ and $L$:
\begin{equation}
\mathcal{A}=-i\omega C_{\mathcal{A}} +\omega^2 C_{\mathcal{A}}^2 R_{\mathcal{A}}+\mathcal{O}(\omega^3)\,,
\end{equation}
where
\begin{equation}\label{eq_CA}
C_{\mathcal{A}}=g_\mathcal{A}\int dE \left(-\frac{\partial f_0}{\partial E}\right)(E-E_F)^\lambda \rho(E)
\end{equation}
and
\begin{equation}
R_\mathcal{A} = \frac{r_{\mathcal{A}}\int dE (E-E_F)^\lambda \left(-\frac{\partial f_0}{\partial E}\right)
\text{Tr}[ (S^{\dagger}\partial_E S)^2] }
{\left\{\int dE (E-E_F)^\lambda\left(-\frac{\partial f_0}{\partial E}\right)  \text{Tr} [S^{\dagger}\partial_E S]\right\}^2}\,,
\label{eq_RA}
\end{equation}
where $g_\mathcal{A}$ and $r_\mathcal{A}$ are constants to be specified below.

In Eq. \eqref{eq_CA} we define $\rho(E)=\text{Tr}[S^{\dagger}\partial_E S]/2\pi i$
as the density of states for the capacitor plate \cite{nig06}. Notice that our formalism considers scattering states only.
Thus, contributions to $\rho(E)$ due to bound states are not included. Furthermore, the index $\lambda$
in Eqs.~\eqref{eq_CA} and~\eqref{eq_RA} denotes the type of transport:
$\lambda=0$ (electric), $\lambda=1$ (thermoelectric) or $\lambda=2$ (thermal).
We now discuss their main properties.
\begin{figure}[!h]
\begin{center}
\includegraphics[width=9cm]{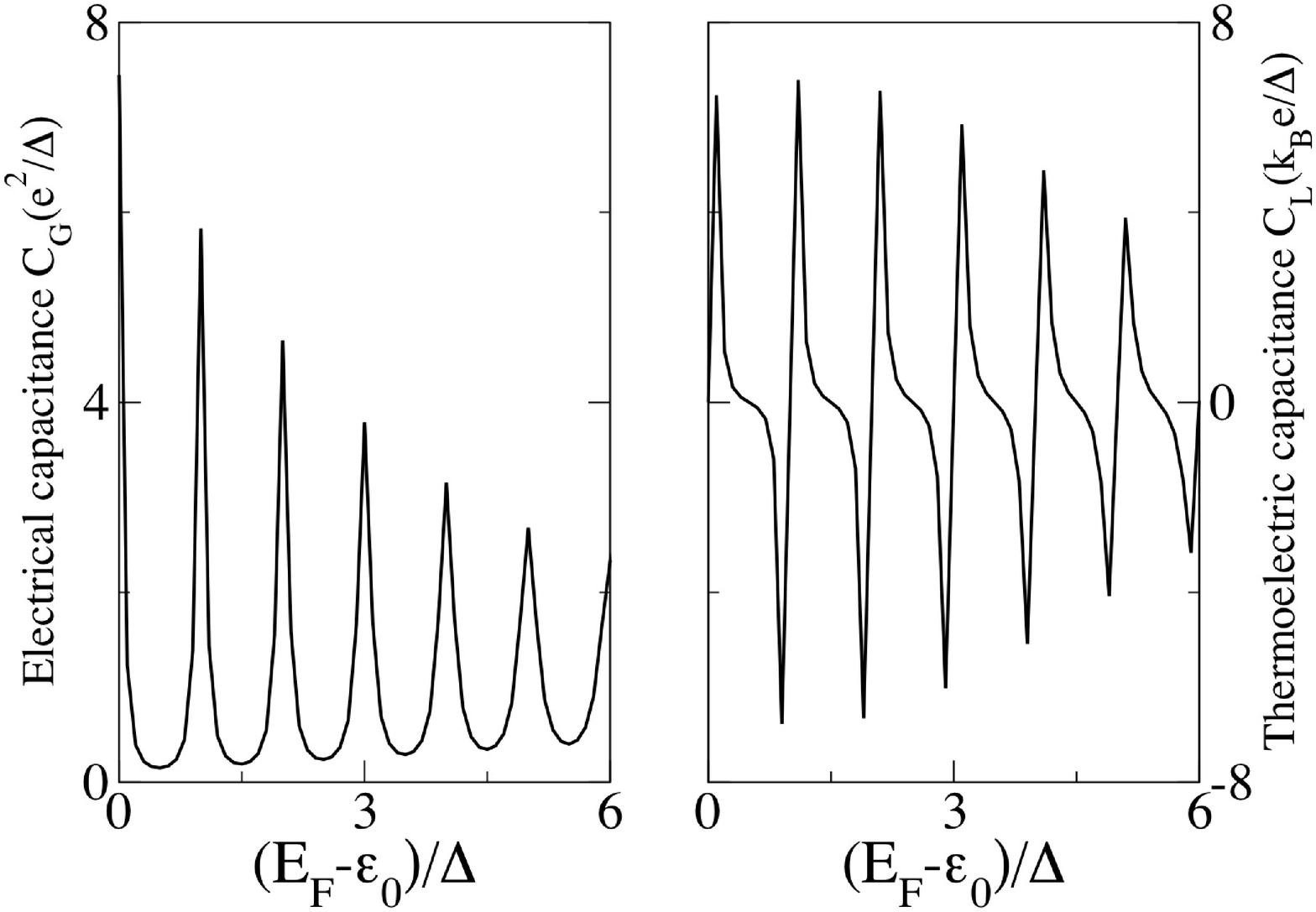}
\caption{(a) Charge and (b) thermoelectric capacitances
as a function of the gate voltage
for the quantum system depicted in Fig.~\ref{figure1}.
Parameters: $\varepsilon_0=\Delta$, $\Omega=3\Delta$, $T_0=\Delta/10$.}
\label{fig1}
\end{center} 
\end{figure}

\begin{figure}[!h]
\begin{center}
\includegraphics[width=8cm]{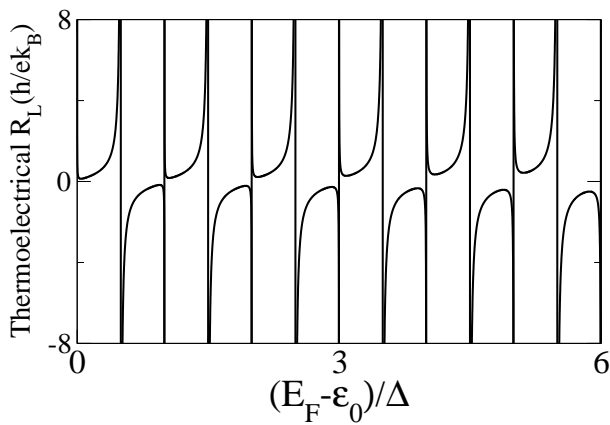}
\caption{Thermoelectrical relaxation resistance as a function of the gate voltage.
We take the same parameters as Fig. \ref{fig1}. }
\label{fig2}
\end{center} 
\end{figure}

\paragraph{Electric admittance.}For $\mathcal{A}=G$ we have $g_G=e^2$.
In this case, $C_{G}$ is a quantum capacitance given by the density of states
and can be accessed experimentally \cite{smi86}. Equations~\eqref{eq_CA} and~\eqref{eq_RA}
are generally valid for an arbitrary $N$-channel conductor.
However,  we now focus on the $N=1$ case because
it is the experimentally relevant situation.
We model the mesoscopic capacitor as a quantum dot coupled to a reservoir~\cite{gab06,feb07}
by means of a single-mode quantum point contact with energy dependent
transmission probability $\mathcal{T}(E)=1/(1+\exp{[-(E-\varepsilon_0)/\Omega}])$, where  $\Omega$ is a characteristic
energy scale of the potential curvature and $\varepsilon_0$ is the transverse energy of the first propagating mode
in the point contact. A strong magnetic field
is applied perpendicular to the dot and transport occurs along edge states.  
Let $\phi=2\pi E/\Delta$ be the phase that an electron picks up after a single turn around the dot, see Fig.~1(a).
Then, the scattering amplitude is~\cite{gab06,feb07} $S(E)=[\sqrt{1-\mathcal{T}}-\exp{(i\phi)}]/[1-\sqrt{1-\mathcal{T}}\exp{(i\phi)}]$,
from which the density of states $\rho(E)$ easily follows. In Fig.~\ref{fig1}(a)
we plot $C_{G}$ as a function of the energy difference $E_F-\varepsilon_0$.  Each peak is associated
with a level in the quantum dot.
Finally, the charge relaxation resistance $R_G$ is obtained from Eq.~\eqref{eq_RA} by setting $\lambda=0$
and $r_G=h/2e^2$. In the single-channel case, one recovers the universal value $R_G=r_G$
valid at sufficiently low temperatures.

\paragraph{Heat admittance.}$K(\omega)=-i\omega C_K+\omega^2C_K^2R_K$ is found from
Eqs.~\eqref{eq_CA} and~\eqref{eq_RA} with $g_K=1/T_0$ and $r_K=hT_0/2$, respectively.
Similarly to the electric case, the imaginary part is related to a capacitance. Here, $C_K$
has the dimensions of a {\em heat capacity}. In fact, $C_K$ is the measured quantity
in a low-frequency ac calorimetry experiment~\cite{bou05}. At low temperature, we find
to leading order in a Sommerfeld expansion $C_K=\pi^2 k_B^2 T_0\rho(E_F)/3$.
Higher orders can be safely neglected for temperatures smaller than the
energy variation of $\rho$, which can be estimated as the mean level spacing $\Delta$.
Therefore, the density of states determines
not only the charge (quantum) capacitance but also its energy counterpart.
More interestingly, the dissipative part of $K(\omega)$ becomes at low temperature
\begin{equation}\label{eq_RK}
R_{K}=\frac{3h}{2\pi^2 k_B^2 T_0}\,,
\end{equation}
This result has a simple interpretation. We recall that $G_{th}=\pi^2 k_B^2 T_0/3h$
is the thermal conductance quantum, recently measured in a suspended nanostructure~\cite{sch00}.
Hence, $R_K=1/(2G_{th})$ is the associated thermal relaxation resistance for a single mode
mesoscopic capacitor. Equation~\eqref{eq_RK} is also universal at low temperature.
We note in passing that the characteristic time $\tau_K=R_K C_K=h\rho(E_F)/2$
is always positive, as it should be, and
equals the $RC$ time for pure electric transport, $\tau_G=R_G C_G=h\rho(E_F)/2$ ~\cite{but93a}.
Furthermore, Eq.~\eqref{eq_RK} scales as $1/N$ with the number of conducting channels, as occurs for the electrical case, because the numerator of Eq.~\eqref{eq_RA} goes as $N$
and its denominator as $N^2$.

\paragraph{Thermoelectric admittance.}Coupled charge and energy transport is governed
at low frequencies by $L(\omega)=-i\omega C_L + \omega^2 C_L^2 R_L$
with $g_L=e/T_0$ and $r_L=hT_0/2e$ in Eqs.~\eqref{eq_CA} and~\eqref{eq_RA} for $\lambda=1$.
At low temperature the thermoelectric capacitance becomes
\begin{equation}\label{eq_CL}
C_L=\frac{\pi^2}{3} e k_B^2 T_0\rho'(E_F)\,.
\end{equation}
Remarkably, $C_L$ can take positive or negative values for a conductor with a density of states
that strongly depends on energy (e.g., a quantum dot with quasi-localized levels). This is in
sharp contrast with the charge capacitance, which takes positive values only, as can be seen
from its low-$T_0$ expansion: $C_G=e^2\rho(E_F)$. As a consequence, while electric
currents lead $\delta V(\omega)$ by $\pi/2$ in a capacitor, the  thermocurrents lead or lag
$\delta T$ depending on the sign of $C_L$. This observation is confirmed
for our dot capacitor, in which we observe 
periodic changes of sign whenever $E_F-\varepsilon_0$ crosses an energy level in the dot, see Fig. 2(b).

In the Seebeck effect, a voltage is generated in response to a temperature shift under the condition $I=0$.
Thus, the Seebeck coefficient is at linear response $S=-L/G$~\cite{book}. We find in the low frequency limit
that $S$ is given by a ratio of capacitances: $S(\omega)=-C_L/C_G+\mathcal{O}(\omega)$.
From our results in Fig.~\ref{fig1} we expect a strong dependence of $S$ as a function of energy
in a mesoscopic capacitor. At low temperature, we find the Mott formula \cite{cut69}
$S=(\pi^2 k_B^2 T_0/3e)\partial_E \ln\rho(E)|_{E=E_F} $.
Note that the same energy dependence is obtained from a study of the Peltier effect
since the Peltier coefficient is simply $\Pi=T_0 S$ by reciprocity~\cite{mol92}.

We find for the thermoelectric relaxation resistance $R_L$ at low temperature:
\begin{equation}\label{eq_RL}
R_L=\frac{3 h}{e \pi^2k_B^2 T_0}\frac{\rho(E)}{\rho'(E)}\Biggr|_{E=E_F}\,,
\end{equation}
which again scales as $1/N$ for a $N$-channel conductor. Unlike the charge relaxation resistance
and $R_K$ in the limit $T_0\to 0$, the resistance $R_L$ is always a nonuniversal
function that depends on the sample details. Surprisingly enough,
we can recast the inverse of Eq.~\eqref{eq_RL} in a more familar form since
$G_L\equiv R_L^{-1}=eG_{th}\partial_E \ln\rho(E)|_{E=E_F}$ 
resembles the Mott formula for the dc thermopower~\cite{cut69}
but applied to a seemingly unrelated quantity.
In Fig.~\ref{fig2} we plot $R_L$ for the mesoscopic capacitor
as a function of the gate voltage. As expected, the thermoelectric
relaxation resistance never reaches a constant value. Note that the divergent
 behavior of $R_L$ in Fig.~\ref{fig2} occurs whenever the  $\rho'(E_F)$ vanishes.
 Obviously the measurable quantity $\tau_L=R_L C_L=h\rho(E_F)$ is always finite and nonzero.
 Interestingly, the time scale for cross charge-energy transport is two times larger
 than purely charge or heat characteristic times. Then, for frequencies
 around $\tau_G^{-1}$, $\tau_L^{-1}$ or $\tau_K^{-1}$ (all of the same order)
 the $\omega^2$ term becomes comparable to the linear-in-$\omega$ term, allowing
 us to probe the characteristic times. For a quantum capacitor, we approximate
 $\rho(E_F)\sim 1/\Delta$ and estimate 
 $\omega\sim 1$~GHz for a typical value $\Delta=10$~$\mu$eV.
 This frequency is not far from
 the scope of present technology
 for both voltage~\cite{gab06} and temperature~\cite{sho13} biases.

\paragraph{Interactions.} A real calculation
of admittance responses requires knowledge of the charge distribution inside
the sample when the conductor is driven out of equilibrium by ac signals~\cite{but93b}.
Thus, we need to discuss screening effects.
As a first approximation, we treat interactions within a mean-field theory.

Let $\delta U_0$ be the internal potential in the conductor away from its equilibrium value.
For definiteness, we assume that $\delta U_0$ is spatially homogeneous
[see Fig.~\ref{figure1}(a) for an example]. The extension to inhomogeneous fields is straightforward~\cite{but93,san13}.
In the presence of interactions, the linear-response current becomes
\begin{eqnarray}\label{current}
\delta I= G(\omega) \delta V +L(\omega) \delta T + \Pi(\omega)\delta U_0\,,
\end{eqnarray}
with $\Pi(\omega)$ the screening function.
Because $\delta I$ is invariant under a global voltage shift, $\Pi(\omega)=-G(\omega)$.
On the other hand, the conductor is coupled to nearby gates via capacitance couplings.
In the homogeneous case, we consider a single electric capacitance $C$.
Then, the ac current is $\delta I=-i\omega C \delta U_0$ and using current conservation
in Eq.~\eqref{current} we find 
\begin{equation}\label{eq_U0}
\delta U_0=\frac{G(\omega)\delta V + L(\omega)\delta T}{G(\omega)-i\omega C}\,.
\end{equation}
We emphasize that the internal potential changes in response not only to voltage variations
but also to temperature shifts~\cite{san13}.

Substituting Eq.~\eqref{eq_U0} into Eq.~\eqref{current} we obtain the electric and thermoelectric
impedances, $\mathcal{G}^{-1}(\omega)=G^{-1}(\omega)+i/\omega C$ and $\mathcal{L}^{-1}(\omega)=L^{-1}(\omega)+iG(\omega)/\omega CL(\omega)$, respectively. Similarly, the interacting ac responses for the heat flow
to oscillatory temperatures and voltages are $\mathcal{K}(\omega)=K(\omega)-L(\omega)M(\omega)/[G(\omega)-i\omega C]$ and
$\mathcal{M}(\omega)=M(\omega)-G(\omega)M(\omega)/(G(\omega)-i\omega C)$, with the latter obeying reciprocity ($\mathcal{M}=T_0 \mathcal{L}$). Importantly, the four admittances constitute a gauge-invariant theory.

Previous expressions can be further simplified in the low frequency limit. We find
that the thermoelectrical response
$\mathcal{L}(\omega)=(-i\omega C_L^\mu + \omega^2 C_L C_L^\mu R_L)$ is expressed
in terms of a renormalized thermoelectric capacitance
$C_L^\mu=C_L C_G^\mu/C_G$, where
the electrochemical capacitance $C_G^\mu$ is given by the geometrical and
quantum capacitances in series: $(C_G^\mu)^{-1}=C^{-1}+C_G^{-1}$.
Finally,  the interacting heat admittance reads
$\mathcal{K}(\omega)=-i \omega C_K^\mu +\omega^2 [C_K^2 R_K-(C_LR_L+C_MR_M)C_LC_M/(C+C_G)]$,
where the renormalized heat capacitance is $C_K^\mu=C_K-C_MC_L/C$. 

\paragraph{Conclusion.} We have presented
a scattering theory for the dynamical response 
of heat and electrical currents to oscillatory electrical and thermal signals
applied to a mesoscopic capacitor.
Our model includes interactions and focuses on the
low-frequency properties of charge and energy transport.
Importantly, we have found that the off-diagonal admittance matrix elements
show crucial differences
with the pure electric or thermal conductances:
positive or negative phase delays and sample-dependent
relaxation resistances. Our results are relevant both
for thermodynamic characterizations of nanosystems
and for prospect applications of thermoelectric nanodevices
operating in the time domain. 

\paragraph{Acknowledgements.}
We thank M. B\"uttiker and J. Splettstoesser for useful comments.
This work was supported by MINECO Grants No.~FIS2011-23526
and No. CSD2007-00042 (CPAN).

\end{document}